\documentclass[12pt]{article}
\usepackage{amssymb,amsmath,epsfig}
\usepackage{cite}
\allowdisplaybreaks

\begin{document}
\title{\bf Evolution of Compact Stars and Dark Dynamical Variables}
\author{M. Z. Bhatti$^1$ \thanks{mzaeem.math@pu.edu.pk}, Z. Yousaf$^1$
\thanks{zeeshan.math@pu.edu.pk} and
M. Ilyas$^2$ \thanks{ilyas\_mia@yahoo.com},\\
$^1$ Department of Mathematics, University of the Punjab,\\
Quaid-i-Azam Campus, Lahore-54590, Pakistan\\
$^2$ Centre for High Energy Physics, University of the Punjab,\\
Quaid-i-Azam Campus, Lahore-54590, Pakistan}

\date{}
\maketitle
\begin{abstract}
This work is aimed to explore the dark dynamical effects of $f(R,T)$
modified gravity theory on the dynamics of compact celestial star.
We have taken the interior geometry as spherical star which is
filled with imperfect fluid distribution. The modified field
equations are explored by taking a particular form of $f(R,T)$
model, i.e., $f(R,T)=f_1(R)+f_2(R)f_3(T)$. These equations are then
utilized to formulate the well-known structure scalars under the
dark dynamical effects of this higher order gravity theory. Also,
the evolution equations for expansion and shear are formulated with
the help of these scalar variables. Further, all this analysis have
been made under the condition of constant $R$ and $T$. We found a
crucial significance of dark source terms and dynamical variables on
the evolution and density inhomogeneity of compact objects.
\end{abstract}

\section{Introduction}

After many observational astronomical consequences coming from
Supernovae Type Ia, BICEP, and cosmic microwave background radiation
\cite{c1a, c1b, c1c}, it has been affirmed that our cosmos is in
accelerated expanding phase. Dark energy (DE) is thought to be
reliable source behind this enigmatic behavior of the universe. In
order to deal with its nature, the modified gravity theories are
believed to be one of the mathematical tools. These theories are
explored by modifying the Einstein-Hilbert (EH) action and
extensively been applied to study the nature and problem of DE, that
may lead to confront with accelerating cosmic expansion \cite{v41,
b2a, R-DE-MG,martin1}.

Nojiri and Odintsov \cite{ya3} introduced some modified
gravitational models for the complete description of early and
late-time evolutionary universe eras. The simplest generalization of
GR includes $f(R)$ gravity in which $R$ is the Ricci scalar. Harko
\textit{et al.} \cite{ya9} modified this theory by invoking
corrections coming from the trace of energy-momentum tensor ($T$) in
the EH action. This theory is widely known as $f(R,T)$ gravity
theory. The motivation of this theory stems from the fact that the
influences introduced by quantum effects or dark non-ideal
relativistic matter environment are being invoked in this analysis.

Houndjo \cite{ya10} obtained some observationally well-consistent
$f(R,T)$ models that could assist enough to analyze the behavior of
matter dominated universe epochs. Baffou \textit{et al.} \cite{ya13}
used perturbation scheme to investigate the viability of few cosmic
models by taking de-Sitter and power law formulations. Bamba
\emph{et al.} \cite{z5d} checked the role of dark source terms
coming from modified gravity models on the dynamics of accelerating
and expanding universe. Durrer and Maartens \cite{6} presented some
results to the credibility of GR in terms of $f(R)$ gravity.
Gravitational stabilities of relativistic compact structures were
examined in in $f(R)$ gravity by \cite{mart5,mart6,mart7}. Yousaf
and his colleagues examined the rate of gravitational implosion with
the help of various modified gravity models for the planar
\cite{7p}, spherical \cite{7s} as well as cylindrical \cite{7c}
relativistic objects. Moraes \emph{et al.} \cite{9} calculated
modified hydrostatic expression for investigating the dynamical
features of some strange and neutron stars with the help of
$f(R,T)=R+2\lambda T$ model.

Herrera \emph{et al.} \cite{12,13} discussed the gravitational
implosion of cylindrical as well as spherical collapse via some
specific boundary conditions. Tewari \emph{et al.} \cite{15}
investigated spherical anisotropic collapse and presented new class
of relativistic models that could be helpful to understand various
dynamical features of stellar models. Sharif and Bhatti \cite{bhat1}
examined the role of adiabatic index as well as physical parameters
on the onset of gravitational collapse of axially symmetric
self-gravitating systems. Sharif and Yousaf \cite{16rt,16r}
considered the problem of dynamical instability of celestial bodies
in modified gravity and found the role of $f(R,T)$ and $f(R)$ in the
subsequent phases of collapsing systems. Recently, Yousaf \emph{et
al.} \cite{epjc1,cqg} joined interior non-static anisotropic
cylindrical system with exterior Einstein-Rosen bridge and
investigated the impact of modified corrections on the onset of
dynamical instability.

The inhomogeneous state is found to be the predecessor in the
process of gravitational collapse for the initially homogenous
stellar structures. It is pertinent to mention that one can
understand some dynamical properties of self-gravitating systems
through investigating the behavior of pressure anisotropy, tidal
forces, inhomogeneous energy density (IED), etc. There has been
extensive work related to check the cause of IED over the surface of
regular compact objects. The work of Penrose and Hawking \cite{ya26}
is among pioneers works in this direction. They found Weyl tensor as
a key figure in the emergence of IED in the evolution of spherically
symmetric objects. Herrera \textit{et al.} \cite{ya27} calculated
some factors responsible for creating IED over the anisotropic
stellar spheres and inferred that pressure anisotropy may lead the
system to develop naked singularity (NS). Virbhadra \emph{et al.}
\cite{vir1,vir2} provided a mathematical platform under which one
can differentiate between the formation of NS and black holes.

Herrera \textit{et al.} \cite{ya29} described gravitational arrow of
time for the dissipative compact systems by making a relation among
Weyl invariant, pressure anisotropy and IED. Herrera \textit{et al.}
\cite{ya30} examined the influences of IED on the expressions of
shear and expansion evolutions in the presence of electromagnetic
field. Yousaf \emph{et al.} \cite{y1t} covered this problem for
spherical radiating geometries in modified gravitational theory and
concluded that a special combination of $f(R,T)$ gravity model could
significantly interfere in the appearance of IED. Bhatti and his
colleagues \cite{b1t} looked into the reasons behind the maintenance
of IED against gravitational collapse of relativistic interiors in
modified gravity. Herrera \emph{et al.} \cite{ltb} and Herrera
\cite{entropy1} considered the case of non-comoving coordinate
system and checked the reasons for the start up of the spherical
collapse by evaluating transport equations. Yousaf \emph{et al.}
\cite{y2t} modified these results by invoking Palatini $f(R)$
corrections. Recently, Herrera \cite{jpc} illustrated the answer to
the question that why observations of tilted congruences notice
dissipative process in stellar interiors which seem to be isentropic
for non-tilted observers.

This paper is a continuation of a previous work presented by
\emph{Herrera et al.} \cite{ya30} in order to check the role of
$f(R,T)=R+\lambda R^2T^2$ cosmic model in the formulations of
structure scalars, shear and expansion evolution equations. The
paper is outlined as under. Next section is devoted to describe some
essential required to understand $f(R,T)$ gravity as well as
spherical distribution of radiating fluids. In section \textbf{3},
we shall compute modified form of structure scalars in the realm of
$R+\lambda R^2T^2$ corrections. We shall also examine the role of
shear and expansion evolution equations in this gravity. Section
\textbf{4} demonstrates the role of scalar parameters in the
emergence of IED of the dust relativistic cloud in today values of
$R$ and $T$. The brief description as well as conclusion are
reported in the last section.

\section{Radiating Sphere and $f(R,T)$ Gravity}

In $f(R,T)$ gravity, the EH action can be written as \cite{ya9}
\begin{equation}\label{1}
A=\int d^4x[f(R,T)+L_M]\sqrt{-g},
\end{equation}
where $g,~T$ are the traces of metric and usual energy-momentum
tensors, respectively, while $L_M$ is the matter Lagrangian. In the
following calculations, we shall opt relativistic units that gives
$8{\pi}G=c=1$. After considering $L_M=\mu$ (where $\mu$ is the fluid
energy density) and applying variations in the above modified action
with $g_{\alpha\beta}$, one can write field equations as
\begin{equation}\label{2}
{G}_{\lambda\nu}={{T}_{\lambda\nu}}^{\textrm{eff}},
\end{equation}
where
\begin{align*}\nonumber
{{T}_{\lambda\nu}}^{\textrm{eff}}&=\left[-\mu
g_{\lambda\nu}f_T(R,T)+T^{(m)}_{\lambda\nu}(1+f_T(R,T))
+\left(f_R(R,T)-\frac{f(R,T)}{R}\right)\frac{R}{2}\right.+\\\nonumber
&\left.+\left({\nabla}_\lambda{\nabla}_
\nu+g_{\lambda\nu}{\Box}\right)f_R(R,T)\right]\frac{1}{f_R(R,T)}.
\end{align*}
In Eq.(\ref{2}), ${G}_{\alpha\beta}$ is the Einstein tensor, while ${{T}_{\lambda\nu}}^{\textrm{eff}}$
is widely known as effective form of energy-momentum tensor. However, $\nabla_\alpha$, $\Box$, $f_T(R,T)$ and $f_R(R,T)$ stand for
covariant derivative, $\nabla_\alpha\nabla^\alpha$,
partial derivatives with $R$ and $T$, respectively.
One can visualize the detailed illustration of the field equations (\ref{2}) from
\cite{ya9} and \cite{y1t}.

Let us consider an irrotational diagonal non-static form of spherically symmetric metric
\begin{equation}\label{3}
ds^2=H^2(t,r)dr^{2}-A^2(t,r)dt^{2}+C^2(d\theta^{2}
+\sin^2\theta{d\phi^2}),
\end{equation}
in which $A,~B$ and $C$ depend on $t$ and $r$. It is assumed that
above geometry is being coupled with radiating shear locally
anisotropic fluid represented by
\begin{equation}\label{4}
T_{\lambda\nu}={\mu}V_\lambda V_\nu+P_{\bot}h_{\lambda\nu}+\Pi
\chi_\lambda\chi_\nu-2{\eta}{\sigma}_{\lambda\nu}+{\varepsilon}l_\lambda l_\nu+q(\chi_\nu
V_\lambda+\chi_\lambda V_\nu),
\end{equation}
where $\varepsilon$ is radiation density, $q_{\beta}$ is heat flux,
$\Pi\equiv P_r-P_{\perp}$, $h_{\alpha\beta},~\sigma_{\alpha\beta}$
are projection and shear tensor, $P_\bot,~P_r$ are tangential and
radial pressure elements, $\mu$ is the energy density and $\eta$ is
coefficient of shear viscosity. The projection tensor is defined as
$h_{\alpha\beta}=g_{\alpha\beta}+V_{\alpha}V_{\beta}$, while
$\chi^\beta$ and $l^\beta$ are radial and null four-vectors,
respectively. Under co-moving coordinate system, the definitions of
these vectors are found as $V^{\nu}=\frac{1}{A}\delta^{\nu}_{0},~
\chi^{\nu}=\frac{1}{C}\delta^{\nu}_{1},~
l^\nu=\frac{1}{A}\delta^{\nu}_{0}+\frac{1}{B}\delta^{\nu}_{1},~
q^\nu=q(t,r)\chi^{\nu}$. In order to maintain comoving coordinate
frame, these obey relations
\begin{eqnarray*}
&&\chi^{\nu}\chi_{\nu}=1,\quad V^{\nu}V_{\nu}=-1,
\quad\chi^{\nu}V_{\nu}=0,\\\nonumber
&&l^\nu V_\nu=-1, \quad V^\nu q_\nu=0, \quad l^\nu
l_\nu=0.
\end{eqnarray*}
With reference to Eq. (\ref{3}), the shear tensor and scalar
corresponding to expansion tensor are
\begin{equation*}
\sigma A=\left(\frac{\dot{H}}{H}
-\frac{\dot{C}}{C}\right),\quad \Theta A=\left(\frac{\dot{H}}{H}+\frac{2\dot{C}}{C}\right).
\end{equation*}
where overdot describes $\frac{\partial}{\partial t}$.

In order to have observationally well-consistent gravitational
theory, one need to cope appropriate $f(R,T)$ gravity model. In this
perspective, we take the following combinations of $f(R,T)$ model
\cite{ya33}
\begin{equation}\label{5}
f(R,T)=f_1(R)+f_2(R)f_3(T).
\end{equation}
This form of model description states a minimal background of matter
and geometry coupling, thereby indicating higher order corrections
in well-known a $f(R)$ theory. Realistic $f(R,T)$ models can be
achieved by picking any Ricci scalar function from \cite{no1} along
with any linear form of $T$ function. In this context, we shall take
$f(R,T)=R+\lambda R^2T^2$, where $\lambda\ll1$. The dynamics
proposed by Einstein can be found on setting $\lambda=0$ in the
above model. The $f(R,T)$ field equations for
Eqs.(\ref{3})-(\ref{5}) are
\begin{align}\label{6}
G_{00}&=\frac{A^2}{1+2R\lambda T^2}\left[{\mu}+{\varepsilon}+2T\lambda R^2
-\frac{\lambda}{2}T^2R^2+\frac{\varphi_{00}}{A^2}
\right],\\\label{7} G_{01}&=\frac{AH}{1+2R\lambda T^2}\left[-\frac{(1+2T\lambda R^2)}{1+2R\lambda T^2}(q+{\varepsilon})
+\frac{\varphi_{01}}{AH}\right],\\\label{8}
G_{11}&=\frac{H^2}{1+2R\lambda T^2}\left[\mu
2T\lambda R^2+(1+2T\lambda R^2)(P_r+\varepsilon-\frac{4}{3}\eta{\sigma})
+\frac{\lambda}{2}T^2R^2+\frac{\varphi_{11}}{H^2}\right],\\\label{9}
G_{22}&=\frac{C^2}{1+2R\lambda T^2}\left[(1+2T\lambda R^2)({P_{\bot}}+\frac{2}{3}\eta{\sigma})+\mu
2T\lambda R^2
+\frac{\lambda}{2}T^2R^2+\frac{\varphi_{22}}{C^2}\right],
\end{align}
where
\begin{align}\nonumber
\psi_{00}&=2\partial_{tt}f_R+\partial_tf_R\left(-2\frac{\dot{A}}{A}+\frac{\dot{H}}{H}
+2\frac{\dot{C}}{C}\right)+\frac{\partial_rf_R}{H^2}\left(-2AA'+A^2\frac{H'}{H}
-2A^2\frac{C'}{C}\right),\\\nonumber
\psi_{11}&=-\frac{H^2}{A^2}\partial_{tt}f_R
+\frac{\partial_tf_R}{A^2}\left(-2H^2\frac{\dot{C}}{C}+H^2\frac{\dot{A}}{A}-2H\dot{H}
\right)+\partial_{rr}f_R\\\nonumber
&+\left(\frac{A'}{A}+2\frac{C'}{C}
-2\frac{H'}{H}\right)\partial_rf_R,\\\nonumber
\psi_{01}&=-\frac{A'}{A}\partial_tf_R+\partial_t\partial_rf_R
-\frac{\dot{H}}{H}\partial_rf_R,\\\nonumber
\psi_{22}&=-C^2\frac{\partial_{tt}f_R}{A^2}+\frac{C^2}{A^2}\left(
\frac{\dot{A}}{A}-3\frac{\dot{C}}{C}-\frac{\dot{H}}{H}\right)
\partial_tf_R+\frac{C^2}{H^2}\partial_rf_R\left(\frac{A'}{A}+\frac{C'}{C}
-\frac{H'}{H}\right),
\end{align}
while $G_{\gamma\delta}$ are mentioned in \cite{ya30}. Here, prime
indicates $\frac{\partial}{\partial r}$. The relativistic fluid
4-velocity, $U$, can be given as
\begin{eqnarray}\label{10}
U=D_{T}C=\frac{\dot{C}}{A}.
\end{eqnarray}
The spherical mass function via Misner-Sharp formulations can be
re-casted as \cite{ya32}
\begin{equation}\label{11}
m(t,r)=\frac{C}{2}\left(1+\frac{\dot{C}^2}{A^2}
-\frac{C'^2}{H^2}\right).
\end{equation}
The temporal and radial derivatives of the above equation after
using Eqs.(\ref{6})-(\ref{8}) and (\ref{10}) are found as follows
\begin{align}\label{12}
D_T{m}&=\frac{-1}{2(1+2R\lambda T^2)}\left[U\left\{(1+2T\lambda
R^2)(\bar{P}_r-\frac{4}{3}\eta\sigma)+2T\lambda R^2\mu
-\frac{\lambda}{2}R^2T^2\right.\right.\\\nonumber
&\left.\left.+\frac{\varphi_{11}}{H^2}\right\}+E
\left\{\frac{(1+2T\lambda R^2)}{1+2R\lambda
T^2}\bar{q}-\frac{\varphi_{01}}{AH}\right\}\right],\\\nonumber
D_Cm&=\frac{C^2}{2(1+2R\lambda T^2)}\left[\bar{\mu}+2T\lambda
R^2-\frac{\lambda}{2}R^2T^2
+\frac{\varphi_{00}}{A^2}-\frac{U}{E}\left\{\frac{\varphi_{01}}{AH}\right.\right.\\\label{13}
&\left.\left.-\frac{(1+2T\lambda R^2)}{1+2R\lambda
T^2}\bar{q}\right\}\right],
\end{align}
where over bar notation describes $\bar{X}=\varepsilon+X,$ while
$D_{T}=\frac{1}{A} \frac{\partial}{\partial t}$. The second equation
from the above set of equations provides
\begin{align}\nonumber
m&=\frac{1}{2}\int^C_{0}\frac{C^2}{1+2R\lambda T^2}\left[\bar{\mu}+2T\lambda R^2-\frac{\lambda}{2}R^2T^2
+\frac{\varphi_{00}}{A^2}-\frac{U}{E}\left\{\frac{\varphi_{01}}{AH}\right.\right.\\\label{14}
&\left.\left.-\frac{(1+2T\lambda R^2)}{1+2R\lambda T^2}\bar{q}\right\}\right]dC,
\end{align}
where $E\equiv \frac{C'}{H}$, whose value can be written through mass function as
\begin{eqnarray}\label{15}
E\equiv\frac{C'}{H}=\left[1+U^{2}-\frac{2m(t,r)}{C}\right]^{1/2}.
\end{eqnarray}
Equations (\ref{12})-(\ref{15}) yield
\begin{align}\nonumber
\frac{3m}{C^3}&=\frac{3\kappa}{2C^3}\int^r_{0}\left[\bar{\mu}+2T\lambda R^2
-\frac{\lambda}{2}R^2T^2
+\frac{\varphi_{00}}{A^2}+\frac{U}{E}\left\{\frac{(1+2T\lambda R^2)}{1+2R\lambda T^2}\bar{q}\right.\right.\\\label{16} &\left.\left.
-\frac{\varphi_{01}}{AH}\right\}C^2C'\right]dr,
\end{align}
that connects various structural variable elements, like energy
density, mass function, etc with $f(R,T)$ extra curvature terms. It
is well known that in the spherical case, one can decompose the Weyl
tensor into two different tensors, i.e., the magnetic
$H_{\alpha\beta}$ part and the electric $E_{\alpha\beta}$ part.
These two are defined respectively as
\begin{equation*}
H_{\alpha\beta}=\frac{1}{2}\epsilon_{\alpha \gamma
\eta\delta}C^{\eta\delta}_{~~\beta{\rho}}V^\gamma
V^{\rho}=\tilde{C}_{\alpha
\gamma\beta\delta}V^{\gamma}V^\delta=,\quad
E_{\alpha\beta}=C_{\alpha\phi\beta \varphi}V^{\phi}V^{\varphi},
\end{equation*}
where $\epsilon_{\lambda\mu\nu\omega}\equiv\sqrt{-g}\eta_{\lambda\mu\nu\omega}$
with $\eta_{\lambda\mu\nu\omega}$ as a Levi-Civita symbol. The electric component of the Weyl tensor
can be expressed through fluid's 4 vectors as
\begin{equation*}\nonumber
E_{\lambda\nu}=\left[\chi_{\lambda}\chi_{\nu}-\frac{g_{\lambda\nu}}{3}
-\frac{1}{3}V_\lambda V_\nu\right]\mathcal{E},
\end{equation*}
in which $\mathcal{E}$ represents scalar corresponding to the Weyl
tesnor. The value of $\mathcal{E}$ through spherical geometric
variables are found as
\begin{eqnarray}\nonumber
\mathcal{E}&=&-\frac{1}{2C^{2}}+\left[-\frac{\ddot{B}}{B}+
\left(\frac{\dot{C}}{C}+\frac{\dot{A}}{A}\right)\left(\frac{\dot{B}}{B}-\frac{\dot{C}}{C}\right)+\frac{\ddot{C}}{C}\right]
\frac{1}{2A^{2}}\\\label{17}
&-&\left[-\left(\frac{A'}{A}-\frac{C'}{C}\right)\left(\frac{C'}{C}+\frac{B'}{B}\right)
+\frac{C''}{C}-\frac{A''}{A}\right]\frac{1}{2B^{2}}.
\end{eqnarray}
Another way of writing $\mathcal{E}$ with the inclusion of $f(R,T)$
extra curvature terms is
\begin{align}\nonumber
\mathcal{E}&=\frac{1}{2(1+2R\lambda T^2)}\left[\bar{\mu}+2R^2\lambda T-(1+2R^2\lambda T)(\bar{\Pi}-2\eta\sigma)
-\frac{\lambda}{2}T^2R^2
+\frac{\varphi_{00}}{A^2}\right.\\\nonumber
&\left.-\frac{\varphi_{11}}{B^2}
+\frac{\varphi_{22}}{C^2}\right]-\frac{3}{2C^3}
\int^r_{0}\frac{C^2}{1+2R\lambda T^2}
\left[\bar{\mu}+2R^2\lambda T-\frac{\lambda}{2}T^2R^2
+\frac{\varphi_{00}}{A^2}\right.\\\label{18}
&\left.+\frac{U}{E}\left\{\frac{(1+2R^2\lambda T)}{1+2R\lambda T^2}\bar{q}
-\frac{\varphi_{01}}{AH}\right\} C^2C'\right]dr,
\end{align}
where the bar over $\Pi$ indicates $\bar{\Pi}=\bar{P}_r-P_\perp$.

\section{Modified Scalar Variables and $f(R,T)$ Gravity}

Here, we shall compute structure scalars corresponding to radiating
spherical bodies in $R+\lambda R^2T^2$ gravity. In this background,
we would use two well-known tensors, i.e., $X_{\alpha\beta}$ and
$Y_{\alpha\beta}$. These tensors were proposed by Bel \cite{ya35}
and Herrera \textit{et al.} \cite{ya29,ya30} after orthogonal
splitting of Riemann curvature tensor. These are
\begin{equation}\label{19}
X_{\alpha\beta}=~^{*}R^{*}_{\alpha\gamma\beta\delta}V^{\gamma}V^{\delta}=
\frac{1}{2}\eta^{\varepsilon\rho}_{~~\alpha\gamma}R^{*}_{\epsilon
\rho\beta\delta}V^{\gamma}V^{\delta},\quad
Y_{\alpha\beta}=R_{\alpha\gamma\beta\delta}V^{\gamma}V^{\delta},
\end{equation}
where steric on the right, left and both sides of the tensor
describe operation related to right, left and double dual of that
term, respectively. These tensors with the help of 4-vector
$V_{\alpha}$ and projection tensor, $h_{\alpha\beta}$, can be
written as
\begin{align}\label{20}
X_{\alpha\beta}&=\frac{1}{3}X_{T}h_{\alpha\beta}
+X_{TF}\left(\chi_{\alpha}\chi_{\beta}-\frac{1}{3}h_{\alpha\beta}
\right),\\\label{21}
Y_{\alpha\beta}&=\frac{1}{3}Y_{T}h_{\alpha\beta}+Y_{TF}\left(
\chi_{\alpha}\chi_{\beta} -\frac{1}{3}h_{\alpha\beta}\right),
\end{align}
here $X_T$ and $Y_T$ indicate trace parts of the tensors
$X_{\alpha\beta}$ and $Y_{\alpha\beta}$, respectively, while
$X_{TF}$ and $Y_{TF}$ stand for the trace-free components of the
tensors $X_{\alpha\beta}$ and $Y_{\alpha\beta}$, respectively (for
details, please see \cite{ya31}). Using Eqs.(\ref{6})-(\ref{10}),
(\ref{20}) and (\ref{21}), we obtain
\begin{align}\label{22}
X_{T}&=\frac{1}{1+2R\lambda T^2}\left\{\bar{\mu}+2R^2\lambda T
+\frac{{\varphi}_{00}}{A^2}+\frac{\lambda}{2}R^2T^2\right\},\\\label{23}
X_{TF}&=-\mathcal{E}-\frac{1}{2(1+2R\lambda T^2)}
\left\{(2R^2\lambda T+1)(-2{\sigma}{\eta}+\bar{\Pi})
-\frac{{\varphi}_{22}}{C^2}+\frac{{\varphi}_{11}}{H^2}\right\},\\\nonumber
Y_{T}&=\frac{1}{2(1+2R\lambda T^2)}\left\{6\mu R^2\lambda T+\bar{\mu}+2R^2\lambda T
+3(1+2R^2\lambda T)\bar{P_r}-2\bar{\Pi}(2R^2\lambda T+1)\right.\\\label{24}
&\left.+\frac{{\varphi}_{00}}{A^2}+\frac{{\varphi}_{11}}{H^2}+\frac{2{\varphi}_{22}}
{C^2}++2{\lambda}T^2R^2\right\},\\\label{25}
Y_{TF}&=\mathcal{E}-\frac{1}{2(1+2R\lambda T^2)}
\left\{(\bar{\Pi}-2{\eta}{\sigma})(2R^2\lambda T+1)
-\frac{{\varphi}_{22}}{C^2}+\frac{{\varphi}_{11}}{H^2}\right\}.
\end{align}
The value of $Y_{TF}$ can be followed from Eqs.(\ref{18}) and
(\ref{25}) as
\begin{align}\nonumber
Y_{TF}&=\frac{1}{2(1+2R\lambda T^2)}\left(\bar{\mu}+2R^2\lambda T-2(1+2R^2\lambda T)(\bar
{\Pi}-4{\eta}{\sigma})+\frac{\lambda}{2}T^2R^2\right.\\\nonumber
&\left.+\frac{{\varphi}_{00}}{A^2}-\frac{2{\varphi}_{11}}{H^2}
+\frac{2{\varphi}_{22}}{C^2}\right)-\frac{3}{2C^3}
\int^r_{0}\frac{C^2}{1+2R\lambda T^2}
\left[\bar{\mu}+2R^2\lambda T\right.\\\label{26}
&\left.-\frac{\lambda}{2}T^2R^2+\frac{{\varphi}_{00}}{A^2}+\frac{U}{E}
\left\{\frac{(1+2R^2\lambda T)}{1+2R\lambda T^2}\bar{q}
-\frac{{\varphi}_{01}}{AH}\right\}C^2C'\right]dr.
\end{align}

One can define few particular collections of fluid and dark source
terms as dagger variables as
\begin{align*}
\mu^{\dag}&\equiv\bar{\mu}+2R^2\lambda T+\frac{{\varphi}_{00}}{A^2}, \quad
P^{\dag}_{r}\equiv
\bar{P_r}+\frac{{\varphi}_{11}}{H^2}-\frac{4}{3}{\eta}{\sigma},\\
P^{\dag}_{\bot}&\equiv P_{\bot}+\frac{{\varphi}_{22}}{C^2}+\frac{2}{3}{\eta}\sigma,\\
\Pi^{\dag}&\equiv P^{\dag}_{r}-P^{\dag}_{\bot}=\Pi-2{\eta}{\sigma}
-\frac{\varphi_{22}}{C^2}+\frac{\varphi_{11}}{H^2}.
\end{align*}
In this context, it follows from Eqs.(\ref{22})-(\ref{25}) that
\begin{align}\nonumber
X_{TF}&=\frac{3\kappa}{2C^3} \int^r_{0}\left[\frac{1}{\{1+2R\lambda T^2\}}\left\{\mu^{\dag}
-\frac{\lambda}{2}T^2R^2
+\left(\hat{q}-\frac{\varphi_q}{AB}\right)\frac{U}{E}\right
\}\right.\\\label{27} &\left.\times
C^2C'\right]dr-\frac{1}{2\{1+2R\lambda T^2\}}\left[\mu^{\dag}-\frac{\lambda}{2}T^2R^2\right],\\\nonumber
Y_{TF}&=\frac{1}{2(1+2R\lambda T^2)}\left[\mu^{\dag}-\frac{\lambda}{2}T^2R^2-
2(1+2R^2\lambda T)\Pi^{\dag}+4R^2\lambda T\right.\\\nonumber
&\left.\times\left(\frac{{\varphi}_{11}}{H^2}
-\frac{{\varphi}_{22}}{C^2}\right)\right]-\frac{3}{2C^3}\int^r_{0}\left[\frac{1}{\{1+2R\lambda T^2\}}
\left\{\mu^{\dag}\right.\right.\\\label{28}
&\left.\left.-\frac{\lambda}{2}T^2R^2+\left(\hat{q}-\frac{\varphi_q}
{AH}\right)\frac{U}{E}\right\}C^2C'\right]dr,\\\nonumber
Y_{T}&=\frac{1}{2(1+2R\lambda T^2)}\left[(1+6R^2\lambda T)\mu^{\dag}
-6\varepsilon R^2\lambda T+3(1+2R^2\lambda T)P_r^{\dag}-2(1+2R^2\lambda T)\Pi^{\dag}
\right.\\\label{29} &\left. -2R^2\lambda T\left(\frac{\varphi_{11}}{H^2}+3
\frac{\varphi_{00}}{A^2}\right)+2(2+2R^2\lambda T)\frac{\varphi_{22}}{C^2}-2{\lambda}T^2R^2\right],
\\\label{30}
X_{T}&=\frac{1}{(1+2R\lambda T^2)}\left[\mu^{\dag}-\frac{\lambda}{2}T^2R^2\right].
\end{align}
The GR structure scalars \cite{ya30} can be retrieved by taking
$f(R,T)=R$ in the above equations. These quantities have utmost
relevance in the study of some important dynamical features of
self-gravitating objects, for instance IED, quantity of matter
content. In order to understand the the role of $f(R,T)$ terms on
the shear and expansion evolution of radiating relativistic
interiors, we shall like to compute Raychaudhuri equations. These
relations were also evaluated independently by Landau) \cite{ya36}.
With the help of $f(R,T)$ structure scalars, one can write
\begin{equation}\label{31}
-(Y_{T})=\frac{{\Theta}^{2}}{3}+\frac{2}{3}{\sigma}^{
\alpha\beta}{\sigma}_{\alpha\beta}+V^{\alpha}\Theta_{;\alpha}
-a^\alpha_{~;\alpha},
\end{equation}
thereby describing the importance of one of the $f(R,T)$ scalar
functions in the modeling of expansion scalar evolution equation. In
the similar fashion, we shall calculate shear evolution equation as
\begin{equation}\label{32}
Y_{TF}=a^{2}+\chi^{\alpha}a_{;\alpha}-\frac{aC'}{BC}
-\frac{2}{3}{\Theta}\sigma-V^\alpha
\sigma_{;\alpha}-\frac{1}{3}\sigma^{2}.
\end{equation}
It is pertinent to mention that this equation has been expressed
successfully via $f(R,T)$ structure scalar, $Y_{TF}$. Using
Eqs.(\ref{16})-(\ref{16}), one can write the differential equation
\begin{equation}\label{33}
\left[X_{TF}+\frac{\kappa\mu^{\dag}}{2(1+2R\lambda T^2)}\right]'=
-X_{TF}\frac{3C'}{C}+\frac{\kappa(\Theta-\sigma)}{2(1+2R\lambda T^2)}
\left({q}B-\frac{\varphi_q}{A}\right).
\end{equation}
On solving it for $X_{TF}$, one can identify that it is the $X_{TF}$
which is controlling IED of the spherical dissipative celestial
bodies.

\section{Evolution Equations with Constant $R$ and $T$}

In this section, we shall investigate the influences of $R+\lambda
R^2T^2$ corrections on the formulations of shear, expansion and Weyl
evolution equation for the relativistic dust cloud with constant
curvature quantities. In order to represents constant values of $R$
and $T$, we shall use the tilde over the corresponding mathematical
quantities. In this framework, the spherical mass function in the
presence of $\bar{R}+\lambda \tilde{R}^2\tilde{T}^2$ corrections is
found to be
\begin{align}\label{34}
m&=\frac{1}{2\{1+2R\lambda T^2\}} \int^r_{0}(\mu+2T\lambda R^2)
C^2C'dr-\frac{{\lambda}R^2T^2} {2\{1+2R\lambda T^2\}}\int^r_{0} C^2C'dr,
\end{align}
while the Weyl scalar turns out to be
\begin{align}\label{35}
\mathcal{E}&=\frac{1}{2C^3\{1+2R\lambda T^2\}}
\int^r_{0}\mu'C^3dr-\frac{{\lambda}R^2T^2} {4\{1+2R\lambda T^2\}}.
\end{align}
The widely known equation relating spherical mass with radiating structural
parameters can be recasted as
\begin{align}\label{36} \frac{3m}{C^3}&=\frac{1}{2\{1+2R\lambda T^2\}}
\left[\mu+2T\lambda R^2-\frac{1}{C^3}\int^r_{0}\mu'C^3dr\right]+\frac{{\lambda}R^2T^2}
{2\{1+2R\lambda T^2\}}.
\end{align}
The $f(R,T)$ structure scalars with $\bar{R}+\lambda
\tilde{R}^2\tilde{T}^2$ corrections boil down to be
\begin{align}\label{37} \tilde{X}_{T}&=\frac{1}{\{1+2R\lambda
T^2\}}\left[{\mu}+2T\lambda R^2
-\frac{\lambda}{2}R^2T^2\right],\\\label{38}
\tilde{Y}_{TF}&=-\tilde{X}_{TF}=\mathcal{E},\\\label{39}
\tilde{Y}_{T}&=\frac{1}{2\{1+2R\lambda T^2\}}\left[{\mu}+2T\lambda
R^2+6\mu T\lambda R^2 -2{\lambda}R^2T^2\right].
\end{align}
These equations indicate that $X_T,~Y_T$ and $Y_{TF},~X_{TF}$ are
controlling effects induced by fluid energy density and tidal forces
caused by Weyl scalar, respectively in an environment of $f(R,T)$
extra degrees of freedom. An equation describing the evolution of
inhomogeneity factors in the emergence of IED for dust fluid is
\begin{align}\label{40}
&\left[\frac{\mu}{2\{1+2R\lambda T^2\}}
-\frac{{\lambda}T^2R^2}{4\{1+2R\lambda T^2\}}+\tilde{X}_{TF}\right]'
=-\frac{3}{C}\tilde{X}_{TF}C'.
\end{align}
This equation involves $\tilde{X}_{TF}$ that was pointed out to be
inhomogeneity factor in the context of GR. It is seen from the above
equation that $\mu=\mu(t)$ if and only if
$\tilde{X}_{TF}=0=\lambda$. This shows that, even in
$\tilde{R}+\lambda \tilde{R}^2 \tilde{T}^2$ gravity,
$\tilde{X}_{TF}$ is a IED factor. In the famework of non-interacting
particles evolving with constant $R$ and $T$, the shear as well as
expansion evolution equations turn out to be
\begin{align}\nonumber
V^\alpha\Theta_{;\alpha}+\frac{2}{3}\sigma^2 +\frac{\Theta^2}{3}
-a^\alpha_{;\alpha}&=\frac{1}{\{1+2R\lambda T^2\}}\left[{\mu}+2T\lambda R^2
-\frac{\lambda}{2}T^2R^2\right]\\\label{41}
&=-\tilde{Y}_T,\\\label{42}
V^{\alpha}\sigma_{;\alpha}+\frac{\sigma^{2}}{3}+\frac{2}{3}\sigma
\Theta&=-\mathcal{E}=-\tilde{Y}_{TF}.
\end{align}
These equations have been expressed with the help of
$\tilde{Y}_{TF}$ and $\tilde{Y}_T$.

\begin{figure} \centering
\epsfig{file=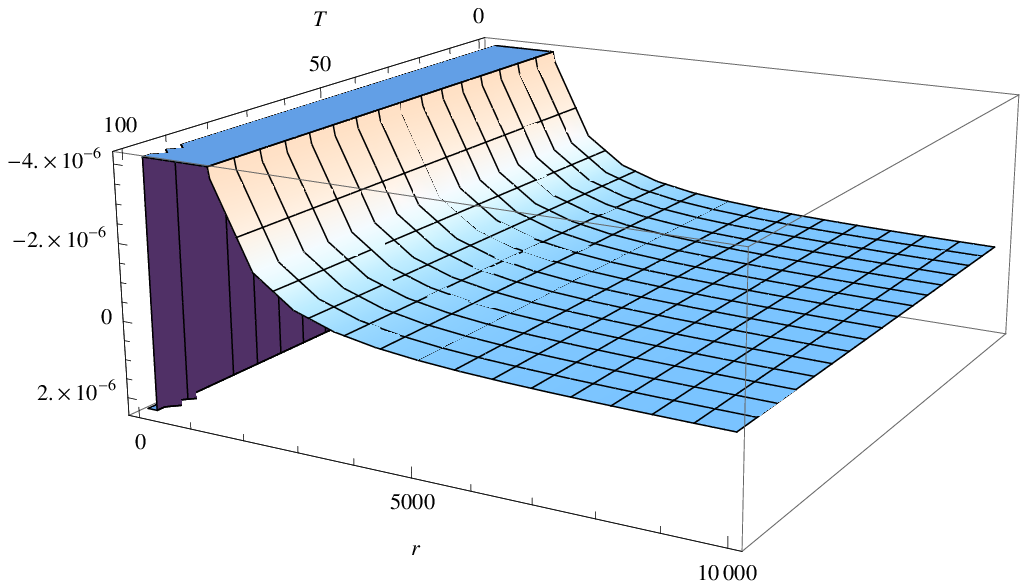,width=1\linewidth} \caption{Plot of the
dynamical variable $\tilde{Y}_{T}$ for the strange star candidate 4U
1820-30.}\label{yt} \epsfig{file=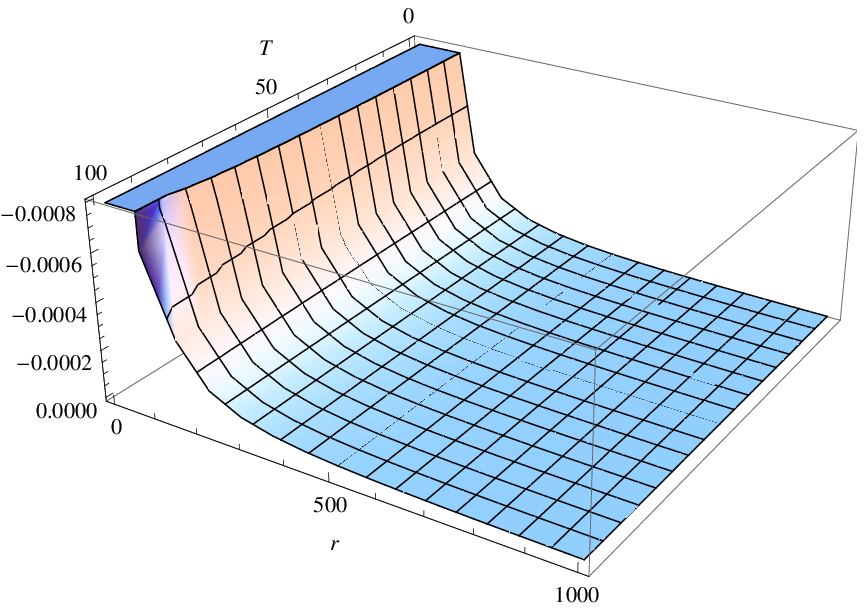,width=1\linewidth}
\caption{Behavior of the dynamical variable $\tilde{X}_T$ for the
strange star candidate 4U 1820-30.}\label{xt}
\end{figure}
\begin{figure} \centering
\epsfig{file=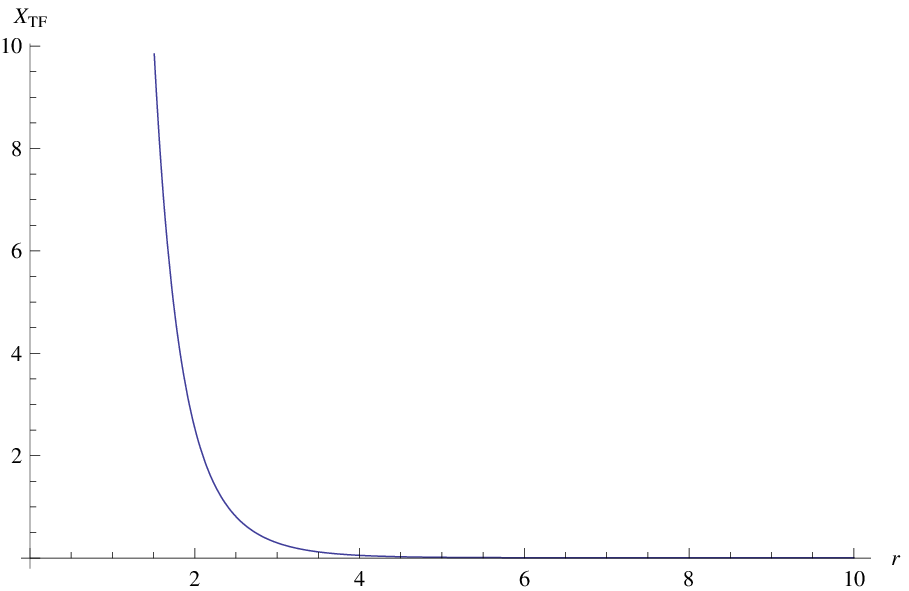,width=1\linewidth} \caption{Role of the dark
dynamical variable $\tilde{X}_{TF}$ on the evolution of the strange
star candidate 4U 1820-30.}\label{xtf}
\epsfig{file=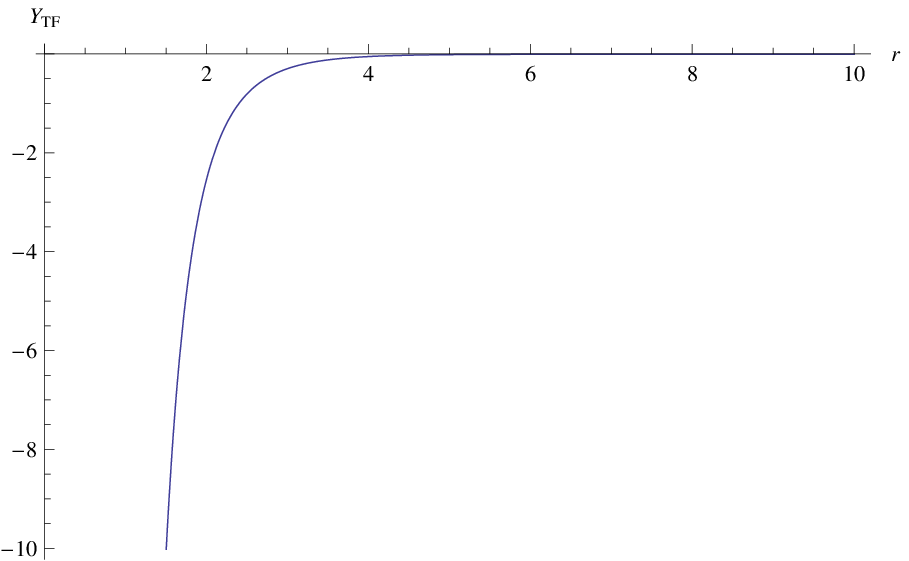,width=1\linewidth} \caption{Plot for the dark
dynamical variable $\tilde{Y}_{TF}$ on the evolution of the strange
star candidate 4U 1820-30.}\label{ytf}
\end{figure}
The study of compact objects is amongst the most burning issues of
our mysterious dark universe in which, stars came into being during
the dying phenomenon of relativistic massive stars. Such celestial
bodies are having size as a big city and generally contain mass
atleast 40\% more mass than solar mass. Due to this fact, their core
density exceeds the density of an atomic nucleus. This specifies
that the compact stars could be treated as test particle to study
some physical features beyond nuclear density.

Rossi X-ray Timing Explorer gathered information based on satellite
observations about the structure of a neutron star, named 4U1820-30.
They found mass of this star to be $2.25M_{\odot}$ containing high
amount of exotic matter. We now apply our results of dynamical dark
variables on the observational values of this compact star. As our
$f(R,T)$ field equations are non-linear in nature, therefore we
suppose that our star consists of non-interacting particles. We
suppose that our geometry is demarcated with the three-dimensional
boundary surface. The interior to that is given by (\ref{3}), while
the exterior vacuum geometry is given by
\begin{equation}\label{ext2}
ds^2_+=-Z^2d\nu^2+Z^{-1}d\rho^2
+\rho^2(d\theta^2+\sin^2{\theta}d\phi^2).
\end{equation}
where $Z=\left(1-\frac{2M}{\rho}\right)$ with $M$ and $\nu$ are
total matter content and retarded time, respectively. We use Darmois
junction conditions \cite{y4} to make continuous connections between
Eqs.(\ref{3}) and (\ref{ext2}) over hypersurface. These conditions,
after some manipulations, provide
\begin{align}\label{ext36}
&Adt\overset{\Sigma}=d\nu\left(1-2\frac{M}{\rho}\right),\quad
C\overset{\Sigma}=\rho(\nu),\\\label{ext37} &
M\overset{\Sigma}=m(t,r),
\end{align}
These constraints should be fulfilled by both manifolds in order to
remove jumps over the boundary.

It is well-known from the literature that the dynamical variable,
$Y_T$ has the same role as that of the Tolman mass density in the
evolutionary phases of those relativistic systems which are in the
state of equilibrium or quasi-equilibrium. Figures (\ref{yt}) and
(\ref{xt}) state the evolution of $\tilde{Y_T}$ and $\tilde{X_T}$
variables with the increase of $r$ and $T$, respectively. Other very
important dark scalar functions are $\tilde{X_{TF}}$ and
$\tilde{Y_{TF}}$. These two variables have opposite behaviors on the
dynamical phases of our relativistic 4U 1820-30 star candidate. The
modified structure scalar $\tilde{X_{TF}}$ is controlling appearance
of inhomogeneities on the initially regular compact object. It can
be observed from the Figure (\ref{xtf}) that the inhomogeneity of
the compact star keep on decreasing by increasing the radial
coordinate of the spherical self-gravitating object. The totally
reverse behavior of $Y_{TF}$ can be observed from Figure
(\ref{ytf}).

\section{Conclusions}

This paper is devoted to explore the effects of extra curvature
ingredients of $f(R,T)$ gravity theory on the dynamical variables of
compact spherical star. The matter contents in the stellar interior
are taken to be imperfect due to anisotropic stresses, shear
viscosity and dissipative terms. A particular form of $f(R,T)$
function i.e., $f(R,T)=f_1(R)+f_2(R)f_3(T)$, is utilized to explore
the modified field equations. The Misner-Sharp mass function is
generalized by including the higher curvature ingredients of
$f(R,T)$ theory. We have disintegrated the Weyl tensor, which
describes the distortion in the shapes of celestial objects due to
tidal forces, in to two parts named as its electric and magnetic
parts. The magnetic part vanishes due to the symmetry of spherical
star and all the tidal effects are due to its electric component.

A correspondence between the the scalar component associated with
Weyl tensor with matter variables has been established under the
influence of extra curvature ingredients of $f(R,T)$ theory. In a
similar way, we have extracted the electric part of Riemann tensor
and formulate its second dual tensor. These couple of tensors are
furthers divided into their constituent scalar parts named as
structure scalars. These scalar parts are written in terms of matter
variables with the help of modified field equations and Weyl scalar.
The effects of higher order terms is also found in the formation of
these dynamical equations as obtained in Eqs.(\ref{27})-(\ref{30}).
These structural dynamical equations have enough significance to
discuss the evolution of self-gravitating compact objects. We have
also explored the Raychaudhuri equations for shear and expansion
scalar which are related with some of the structure scalars.
Further, a differential equation is formulated by adopting the
procedure developed by Ellis which have importance in the discussion
of the inhomogeneous density distribution in the universe. One can
find that the irregularities in the matter density can be controlled
via one of the scalar variable. We have also explored the dark
dynamical variables under the condition of constant Ricci invariant
and trace of stress energy tensor. These dark dynamical variables
are effected by the tidal effects coming due to the electric part of
Weyl tensor. Further, the evolution equations for shear and
expansion are formulated in this background and linked with
structure scalars. All our results are consistent with those already
obtained in the literature.

\end{document}